\documentclass{PoS}
\usepackage{subfigure}
\newcommand{\aap}{A\&A}

\newcommand{\aj}{AJ}
\newcommand{\apj}{ApJ}

\newcommand{\apjl}{ApJ}
\newcommand{\apjs}{ApJS}
\newcommand{\araa}{ARA\&A}
\newcommand{\mnras}{MNRAS}

\newcommand{\procspie}{SPIE}

\title{The central kpc of edge-on AGN}

\ShortTitle{The central kpc of edge-on AGN}

\author{\speaker{Semir Smaji\'c}%
        \thanks{Semir Smaji\'c is member of the International Max Planck Research School (IMPRS) for Astronomy and Astrophysics at the MPIfR and the Universities of Bonn and Cologne.}\\
       I. Physikalisches Institut, Universit\"at zu K\"oln, Germany\\
       Max-Planck-Insitut f\"ur Radioastronomie (MPIfR), Bonn, Germany\\
       E-mail: \email{smajic@ph1.uni-koeln.de}}

\author{Sebastian Fischer\\
        I. Physikalisches Institut, Universit\"at zu K\"oln, Germany\\
        E-mail: \email{fischer@ph1.uni-koeln.de}}
        
\author{Jens Zuther\\
        I. Physikalisches Institut, Universit\"at zu K\"oln, Germany\\
        E-mail: \email{zuther@ph1.uni-koeln.de}}
        
\author{Monica Valencia-S.\\
        I. Physikalisches Institut, Universit\"at zu K\"oln, Germany\\
        Max-Planck-Insitut f\"ur Radioastronomie (MPIfR), Bonn, Germany\\
        E-mail: \email{mvalencias@ph1.uni-koeln.de}}
        
\author{Andreas Eckart\\
        I. Physikalisches Institut, Universit\"at zu K\"oln, Germany\\
        Max-Planck-Insitut f\"ur Radioastronomie (MPIfR), Bonn, Germany\\
        E-mail: \email{eckart@ph1.uni-koeln.de}}

\abstract{The NIR is less influenced by dust extinction than optical light. This enables us to look to an extent through dusty regions. In addition, it is sensitive to the mass-dominating stellar population. The combination of NIR imaging and spectroscopy  of the VLT integral field spectrograph SINFONI gives us the opportunity to analyze several emission and absorption lines and to investigate the stellar population and ionization mechanisms over a field of view (FOV) of 4$\times$4 square arcsec. We detect several emission lines ([Si VI], Pa$\alpha$, Br$\gamma$, H$_2$, [Fe II]) and analyze these to disentangle the ionization mechanisms and the kinematics of the gas. We use the absorption features of Si I, CO(6-3) and CO(2-0) and the large wavelength range of H+K-band for a thorough continuum synthesis. We fit a stellar, dust and power-law contribution together with an extinction component to the H+K-band emission. The result presents regions dominated by the different contributors and extinction information over the full FOV.}

\FullConference{Nuclei of Seyfert galaxies and QSOs - Central engine \& conditions of star formation\\
		 November 6-8, 2012\\
		 Max-Planck-Insitut f\"ur Radioastronomie (MPIfR), Bonn, Germany}

\begin{document}

\section{Introduction}
Most galaxies with a significant bulge component harbor a super massive black hole (SMBH) (e.g., \cite{2008ARA&A..46..475H} and references therein) and most likely every galaxy harbors a massive black hole (MBH), i.e. the black hole at the center of the Milky Way. The evolution of SMBHs can be studied best in active galaxies where the SMBH interacts heavily with the host galaxy. This interaction works both ways, on the one hand gas from the inter stellar medium (ISM) feeds the central SMBH via inflow (e.g., nuclear bars and spirals) and on the other hand the active SMBH blows out gas and radiates photons into the ISM via outflows (e.g., jets and shocks). In which way these two phenomena work is not fully understood, i.e. the effect of outflows which can disrupt but also trigger star formation in the ISM \cite{2007MNRAS.382.1415S}.

The Unified Model (UM) explains the different types of active galaxies with regard to their orientation towards the observer \cite{1993ARA&A..31..473A}. There are attempts to extend the UM by adding an evolutionary component to the scenario \cite{2006ApJ...653..137Z}. This evolutionary component is supposed to bring together the orientation phenomena of the UM, Seyfert 1 and Seyfert 2 galaxies, with the non-hidden broad line region (NHBLR) galaxies and narrow line Seyfert 1 (NLS1) galaxies. However, when analyzing the extending models statistically there is the important issue of biases. A topical example for a biased sample are NLS1 galaxies that show several different characteristics which cannot be combined easily into one class of objects seperate from Seyfert 1 galaxies, see more on this in \cite{2013PoS...M.V-S.}. Another possible bias is that of Seyfert 2 galaxies beeing biased by Seyfert 1 galaxies where the host galaxy, not the dusty 'torus', provides the dust screen that prevents us from seeing the BLR hence classifying these galaxies wrongly as Seyfert 2 galaxies. Observations of polarized light can help to look "behind" the torus by detecting the scattered and reflected broad emission lines. This will usually not work with dust obscuration by the host because this dust obscuration often occurs on large galactic scales and far away from the nucleus where the broad emission lines are excited \cite{2012A&A...544A.105S,2012A&AT...27..557A}. The last problem is supported by the finding that Seyfert 2 galaxies tend to be edge-on oriented \cite{2010ApJ...725L.210S,2011MNRAS.414.2148L}. In addition, Seyfert 2 galaxies show very disturbed morphologies similar to HII and starburst galaxies \cite{2004ApJ...616..707H}.

We will present two individual results from our study of a nearby Seyfert 2 sample selected from the Veron-Cetty \& Veron catalogue \cite{2006A&A...455..773V}.

\subsection{The Data}
The data for this work was obtained with SINFONI the Very Large Telescope (VLT) integral field spectrograph \cite{2003SPIE.4841.1548E}. We have spectral and spatial information over the H+K-band ranging from 1.45 $\mu$m to 2.4 $\mu$m at a field of view (FOV) of 4$\times$4 square arcsec centered on the nucleus. The two galaxies discussed here are NGC 7172 and NGC 289 both Seyfert 2 galaxies chosen from our sample.

\begin{figure*}[htbp]
\centering
	\subfigure[NGC 289]{\includegraphics[width=0.33\textwidth]{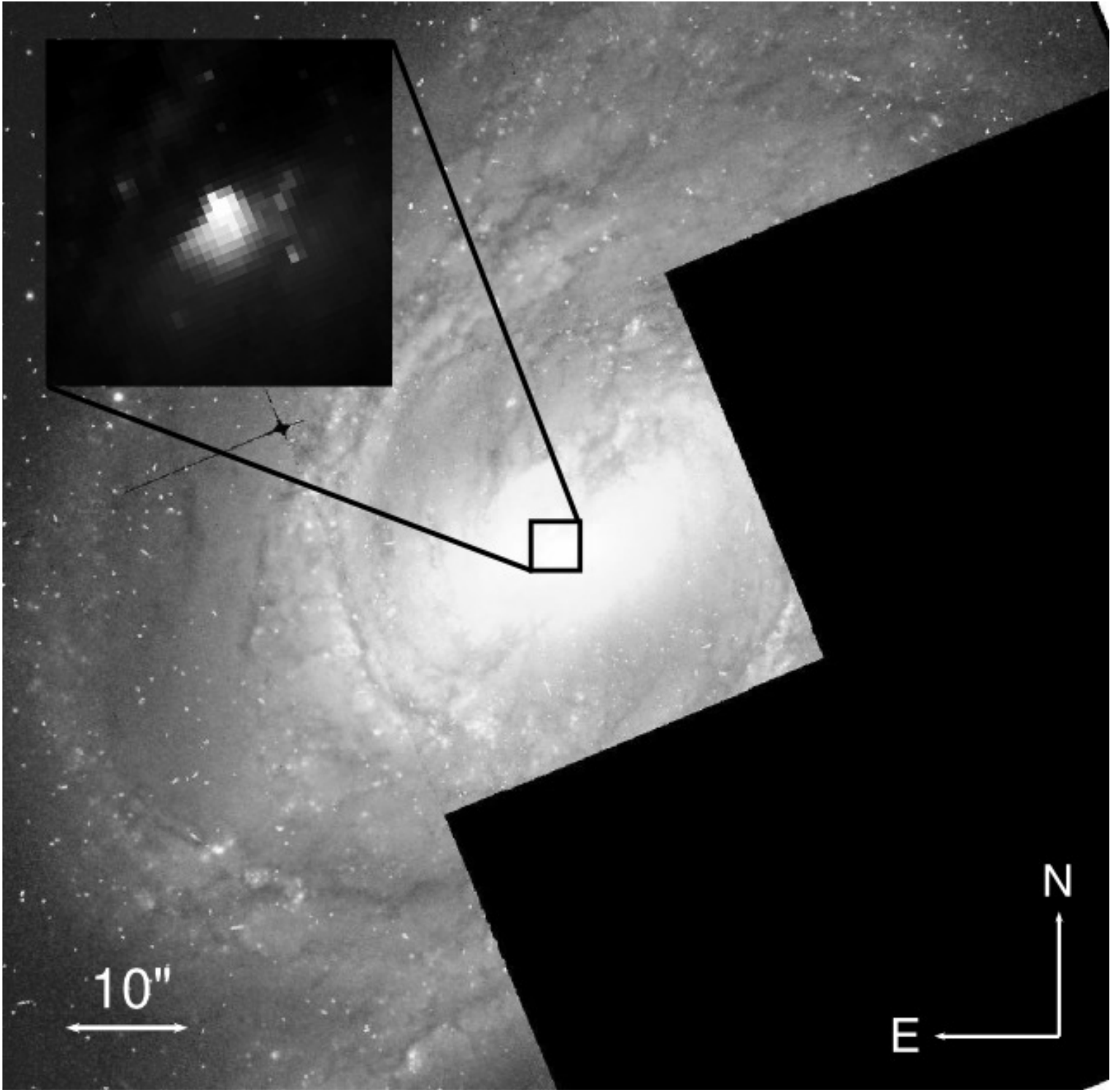}\label{fig:NGC289}}\qquad
	\subfigure[NGC 7172]{\includegraphics[width=0.6\textwidth]{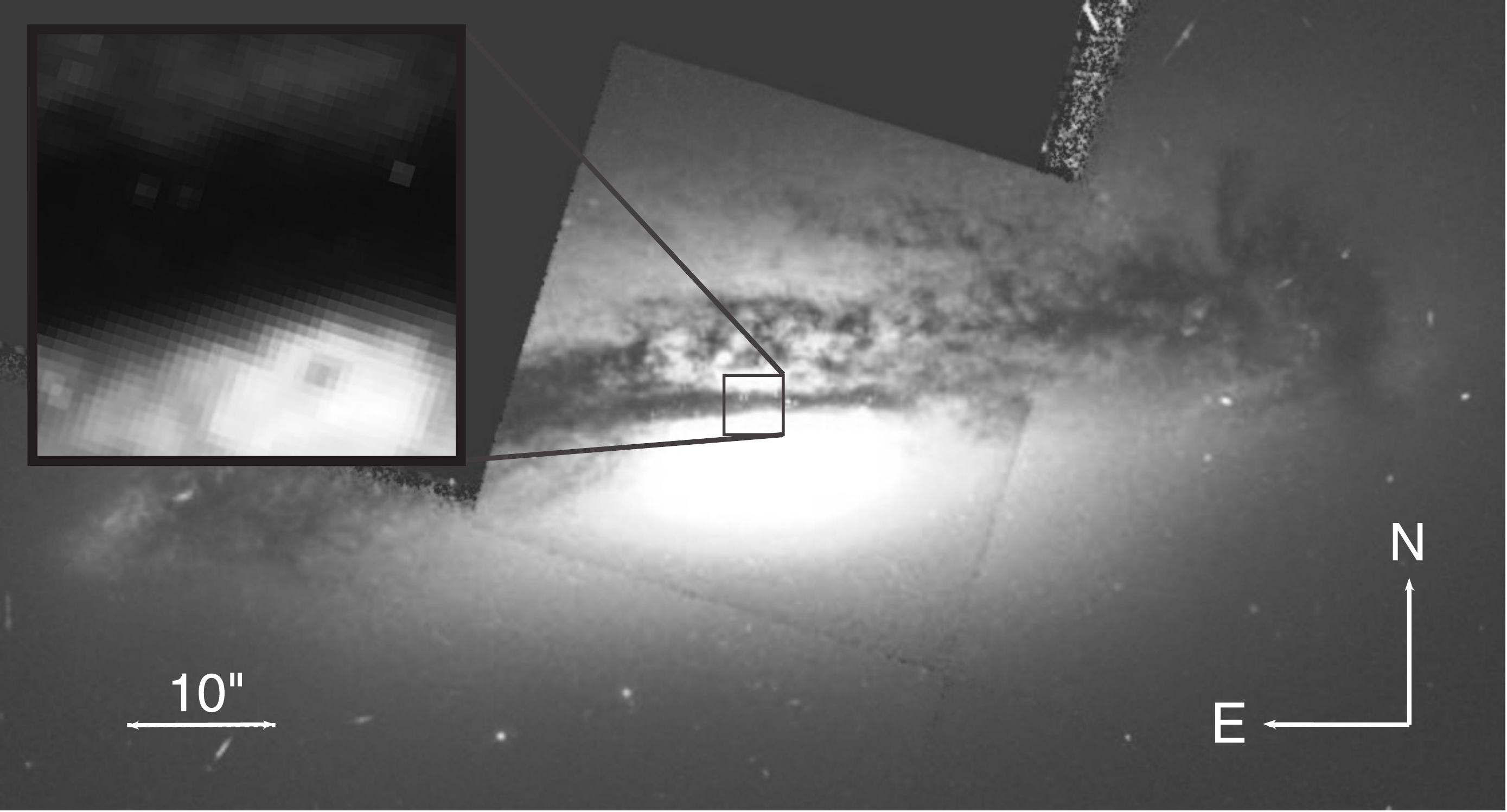}\label{fig:NGC7172}}
 \caption{The images were taken with the HST F606W filter. Both images were overscaled to show the full extent of the host galaxy. In both images the detail in the upper left corner shows the nuclear region using a linear brightness scale. The details show our observations FOV.
}
 \label{fig:HST}
\end{figure*}

NGC 7172 is an edge-on early-type spiral galaxy at a redshift of about 0.0087 and is a member of the Hickson Compact Group 90 (HCG90). It has a prominent dustlane crossing its nuclear region from east to west (see Fig. \ref{fig:NGC7172}). Observations of this galaxy cover a wavelength range from radio over MIR, NIR and optical to X-ray observations. NGC 7172 was identified with the X-ray source H2158-321 \cite{1979ApJS...40..657M}. Polarized observations were not able to detect broad emission lines \cite{2001MNRAS.327..459L}. However, short term variability measured in the X-ray suggests that there is a BLR around the active nucleus of NGC 7172 \cite{1998MNRAS.298..824G}.

NGC 289 is a barred spiral galaxy at a redshift of 0.0054 interacting with its dwarf companion Arp 1981. It has a dustlane crossing its red nuclear region \cite{2002AJ....123..159C} (see also Fig. \ref{fig:NGC289}). A several million years old starburst is supposed to have occurred in the nuclear region of NGC 289 \cite{2004AJ....127.3338B}.

\section{Results}
These two galaxies show several similarities. Both are interacting, NGC 7172 as part of HCG90 and NGC 289 with its dwarf companion. From optical images both galaxies show strong dust patterns close to the nuclear regions and both galaxies do not show strong signs of an active starburst (see Fig. \ref{fig:HST}).

In the following we show that, although the nuclear regions of these galaxies look similar, a closer and better resolved look shows significant differences.

\subsection{Continuum Synthesis}
\label{sec:csynth}
The continuum of galaxies tells us a lot about their properties. From color-color diagrams we are able to determine the amount of stellar light and the extinction of our objects. Since we have additional spectral information we can do a sophisticated and thorough continuum synthesis that will give us more information about the composition of the light emitted by the the nuclear region.

\begin{figure*}[htbp]
\centering
	\subfigure[NGC 289]{\includegraphics[width=0.43\textwidth]{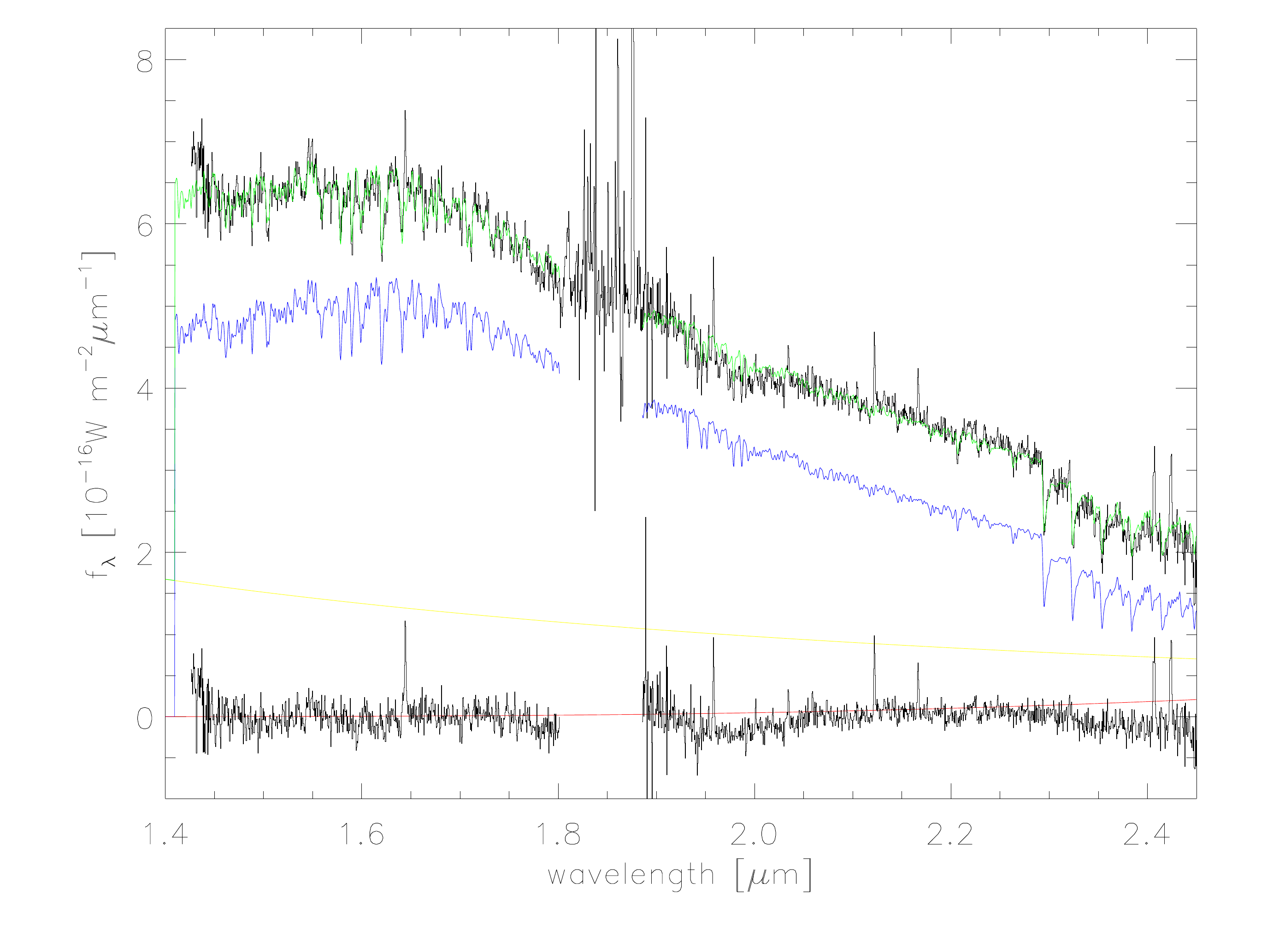}\label{fig:nucspec289}}\qquad
	\subfigure[NGC 7172]{\includegraphics[width=0.43\textwidth]{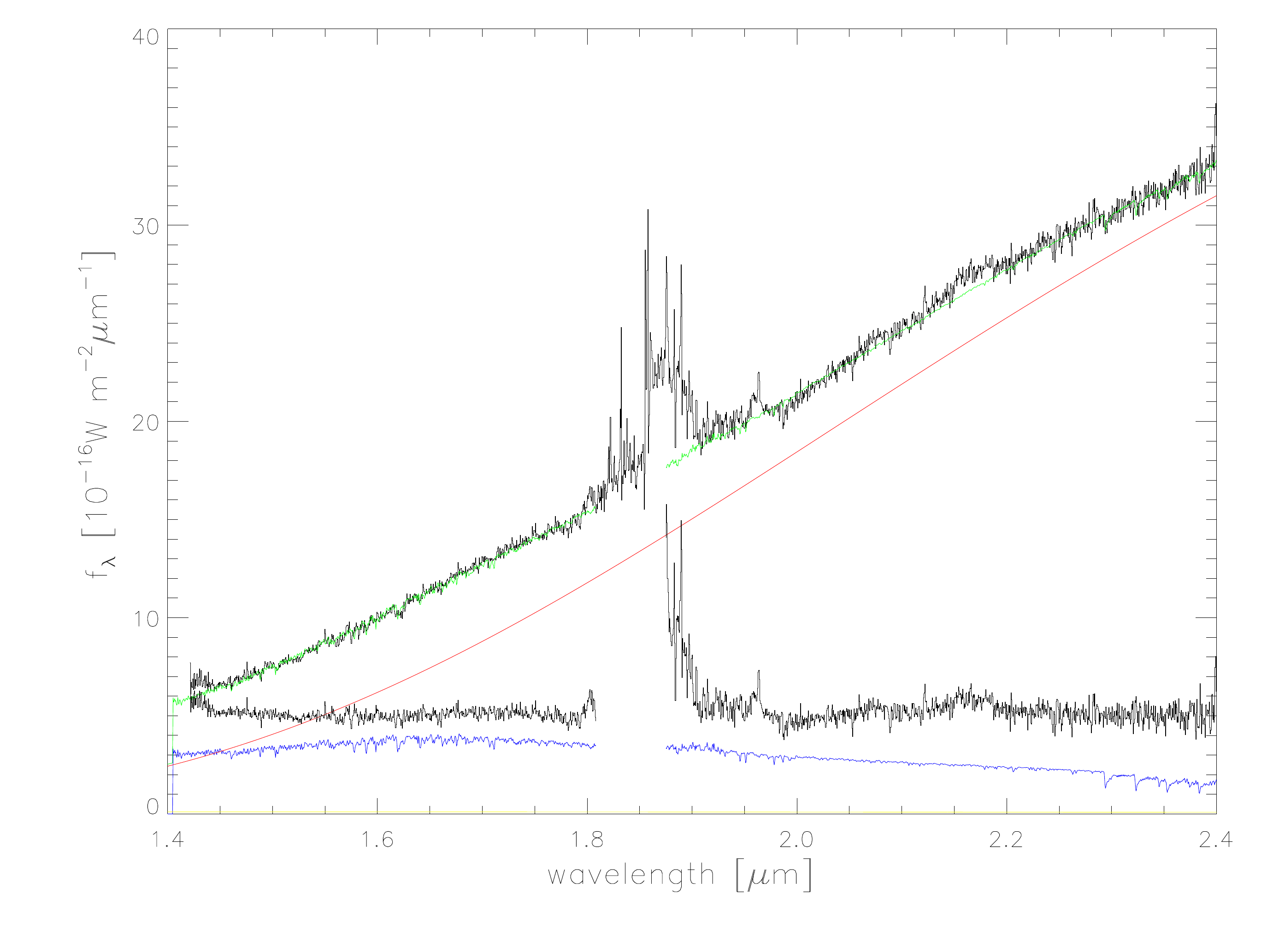}\label{fig:nucspec7172}}
 \caption{The spectra are extracted from the nuclei of NGC 289 and NGC 7172. Overplotted is the continuum synthesis fit (green) and the components: stars (blue), power-law (yellow), dust (red). The flat spectra are the residua. Note that the residuum of NGC 7172 is moved from 0 flux to 5 for displaying reasons.
}
 \label{fig:nucspec}
\end{figure*}

\begin{figure*}[htbp]
\centering
	\subfigure[NGC 289]{\includegraphics[width=0.33\textwidth]{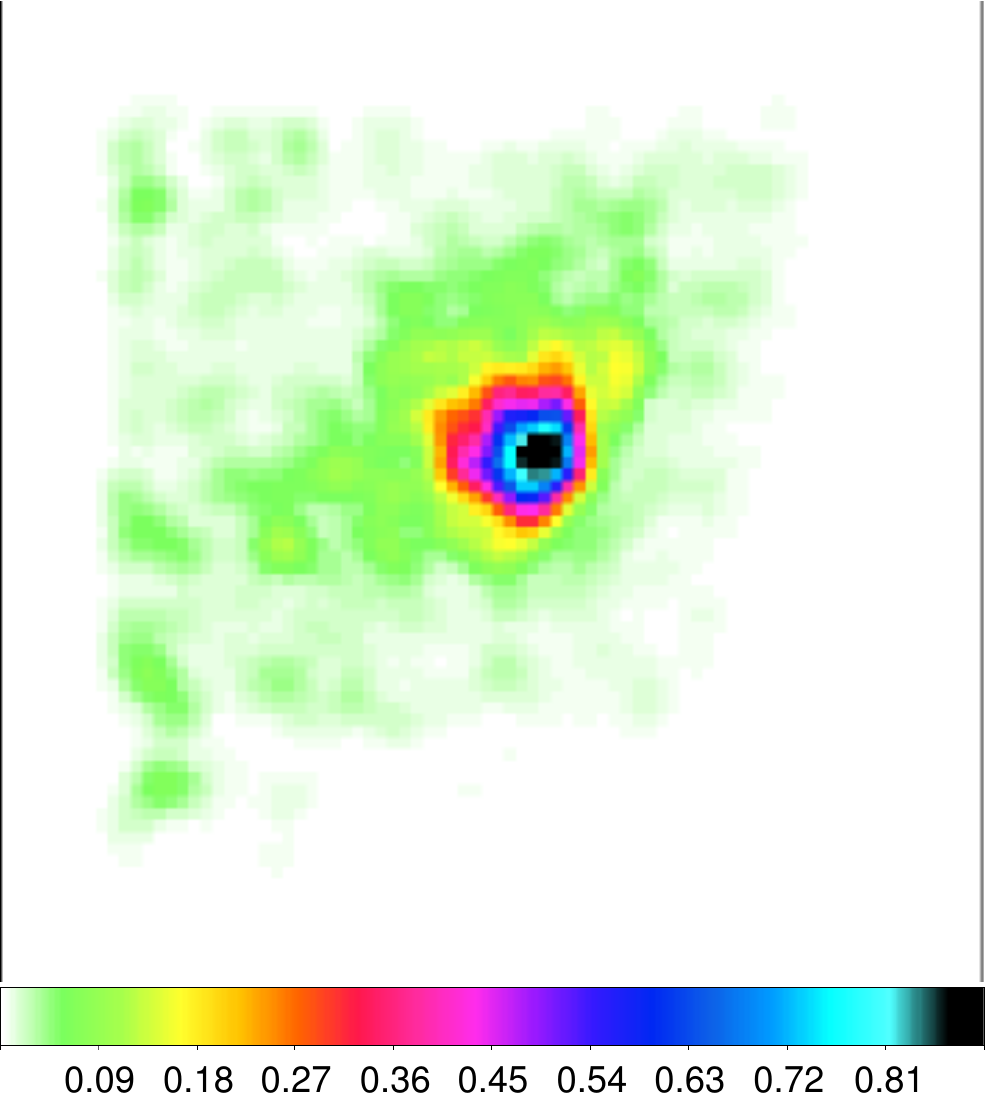}\label{fig:NGC289AV}}\qquad\qquad\qquad
	\subfigure[NGC 7172]{\includegraphics[width=0.33\textwidth]{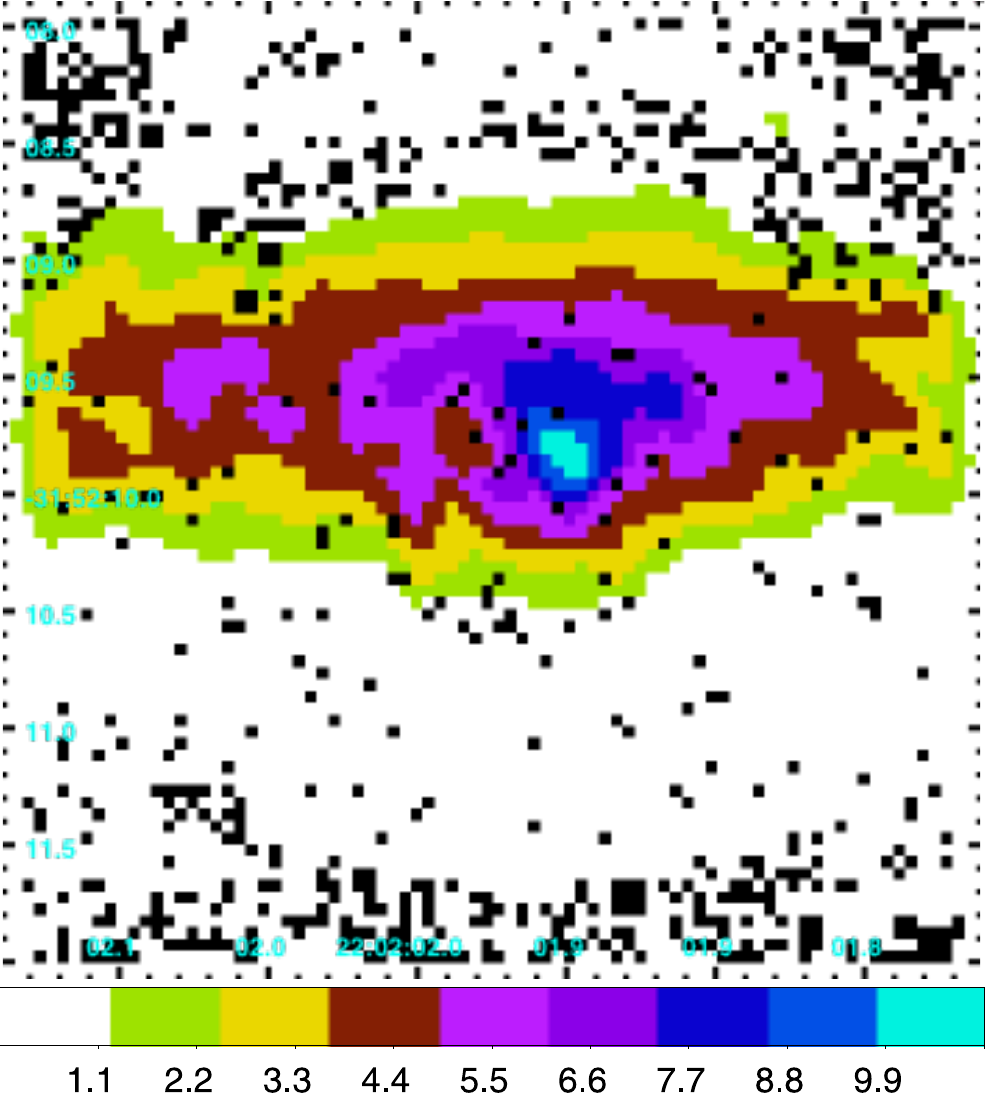}\label{fig:NGC7172AV}}
 \caption{The images show the fitted extinction maps of NGC 289 and NGC 7172. Extinction values are given in A$_V$.
}
 \label{fig:ext}
\end{figure*}

We use our own idl routine with an AMOEBA minimization to get information on the different emitting components that dominate nuclear regions. For our analysis we fit a (hot) dust component, a stellar component and a power-law component multiplied by an extinction component to the spectra of our objects. Figure \ref{fig:nucspec} shows nuclear spectra of our objects overlayed with the different components. Both spectra, the very red spectrum of NGC 7172 and the stellar dominated spectrum of NGC 289, show flat residuals. NGC 289 shows strong stellar absorption features in H-band (e.g., Si I and CO(6-3)) and K-band (e.g., the CO band heads from 2.29$\mu$m on). All these features are fitted well by our stellar template taken from a red supergiant. Hence, the main continuum emission is coming from evolved stars, either from AGB stars or old main sequence stars. The stellar contribution is about 70\% of all the emission in H+K-band. In addition, we fit a power-law component which makes up 30\% of the emission and an almost negligible dust component. NGC 7172 shows a strong red continuum emission which can only be attributed to very hot dust close to the sublimation temperature of about 1200K. The dust emission makes up more than 80\% of the continuum emission on the nucleus. The stellar absorption features are still visible, however, they are filled up and diluted by the strong dust emission. Due to the strong dust component stars are eligible for less than 20\% of all the continuum contribution. However, the total stellar contributions in NGC 7172 and NGC 289 are in the same order of magnitude. The power-law component is negligible for NGC 7172. The extinction component fit worked very well as can be seen in figure \ref{fig:ext}. Our extinction fit traces the dust clouds seen in the Hubble Space Telescope (HST) images \ref{fig:HST}. The extinction values for NGC 289 range up to 1 mag A$_V$. The extinction for NGC 7172 follows exactly the opaque dustlane and reaches an A$_V$ of more than 11 in the central region.

\subsection{Line Emission}
The nuclear spectrum of NGC 289 shows several narrow emission lines. In particular strong molecular hydrogen emission in the K-band (e.g., H$_{2} \lambda$ 1.959, 2.12, 2.23 $\mu$m) and [Fe II] in the H-band. We also detect a rather faint narrow Br$\gamma$ line. The spectrum of NGC 7172 surprisingly shows broad H-recombination lines. Pa$\alpha$, which is situated in the band gap between and H- and K-band, is clearly visible. A fainter broad Br$\gamma$ line is also detected. The full width at half maximum (FWHM) of both broad lines is about 5700 km s$^{-1}$, a typical width for Seyfert 1 galaxies. With the luminosity and the FWHM of the Pa$\alpha$ line we are able to determine a black hole mass of the SMBH \cite{2010ApJ...724..386K}. With a luminosity of L$_{Pa\alpha}$=9.1$\times$10$^{41}$ erg s$^{-1}$ and a FWHM of 5700 km s$^{-1}$ we determine a black hole mass of M$_{BH}$=(4.5$\pm$1.0)$\times$10$^{8}$M$_{\odot}$. This black hole mass is similar to other literature values \cite{1994A&A...282..418L,2006ApJ...645..928A,2009ApJ...690.1322W}. We also detect the narrow emission lines of Pa$\alpha$, Br$\gamma$, [Fe II] $\lambda$1.56 $\mu$m, H$_{2}$ $\lambda$ 1.959, 2.12 $\mu$m and the coronal line [Si VI] $\lambda$1.962 $\mu$m. Due to the detection of the [Si VI] coronal line we are sure that we are looking at the position of the SMBH. Because of the high ionization potential of more than 150 eV for [Si VI] this line can only be excited by the radiation from an accreting SMBH. The coronal line region (CLR) is supposed to be situated between the narrow line region (NLR) and BLR. The occurrence of a CLR is sometimes connected to an outflow in the same direction. The 8 cm VLA map shows an elongation into the south-west similar to that of the [Si VI] line indicating a possible outflow (see Fig. \ref{fig:sivifl}). The flux peak of the H$_{2}$ lines is centered on the position of the SMBH, whereas [Si VI] and Pa$\alpha$ are slightly offset to the south-west. Pa$\alpha$ and H$_{2}\lambda$2.12$\mu$m are also detected along the galactic plane (east-west direction). The line of sight velocity (LOSV) maps of these two lines indicate a rotation from east (blue shifted) to west (red shifted) with velocities of up to 150 km s$^{-1}$. The LOSV dispersion (LOSVD) does not show any specific increase or decrease which would be indicative of an outflow or an inner disks, see Fig. \ref{fig:figures}.

\subsection{Stellar Content}
As already seen in section \ref{sec:csynth} the main contribution of the stellar light comes from evolved stars with typical absorption features (e.g., CO(6-3),CO(2-0)). The problem here is that AGB stars and old stars look very similar with respect to their spectral energy distribution (SED) and absorption features in the NIR. There are still possibilities to identify starburts in the NIR, i.e. the Br$\gamma$ equivalent width (EW) and H-band luminosity. But, one has to be careful as not to confuse the Br$\gamma$ emission origin from a starburst with an NLR origin if the emission is located close to the nucleus. This is exactly the case in the nuclear region of NGC 7172. Hence, we were not able to do a stellar population synthesis. From our continuum synthesis we were still able to get information about the scale of the total stellar emission (section \ref{sec:csynth}), the stellar LOSV and the stellar LOSVD. We show an example of a nuclear spectrum in figure \ref{fig:nucspec}.

\begin{figure*}[htbp]
\centering
	\subfigure[Pa$\alpha$ narrow flux]{\includegraphics[width=0.28\textwidth]{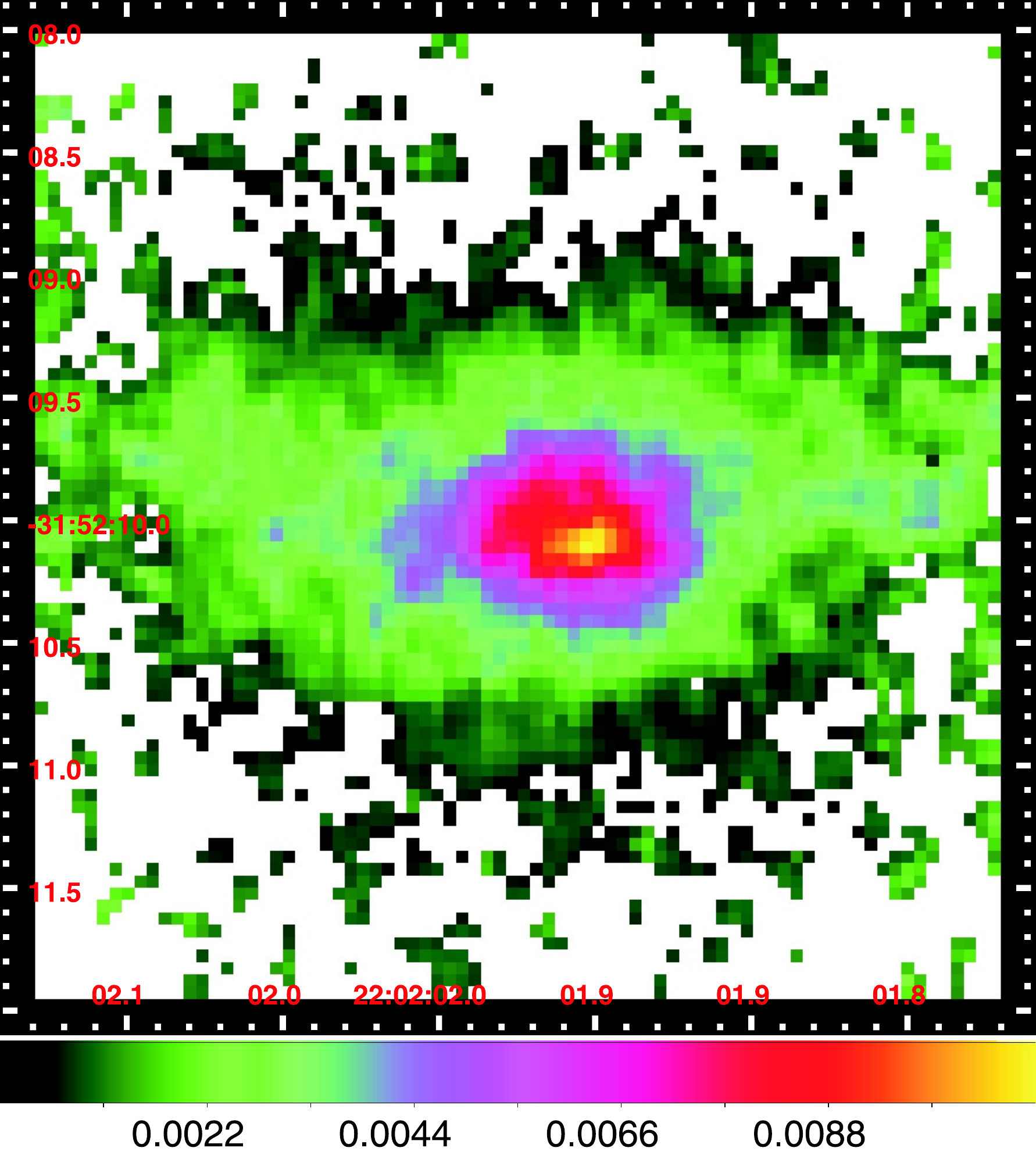}\label{fig:Paafl}}\quad
	\subfigure[Pa$\alpha$ narrow LOSV]{\includegraphics[width=0.28\textwidth]{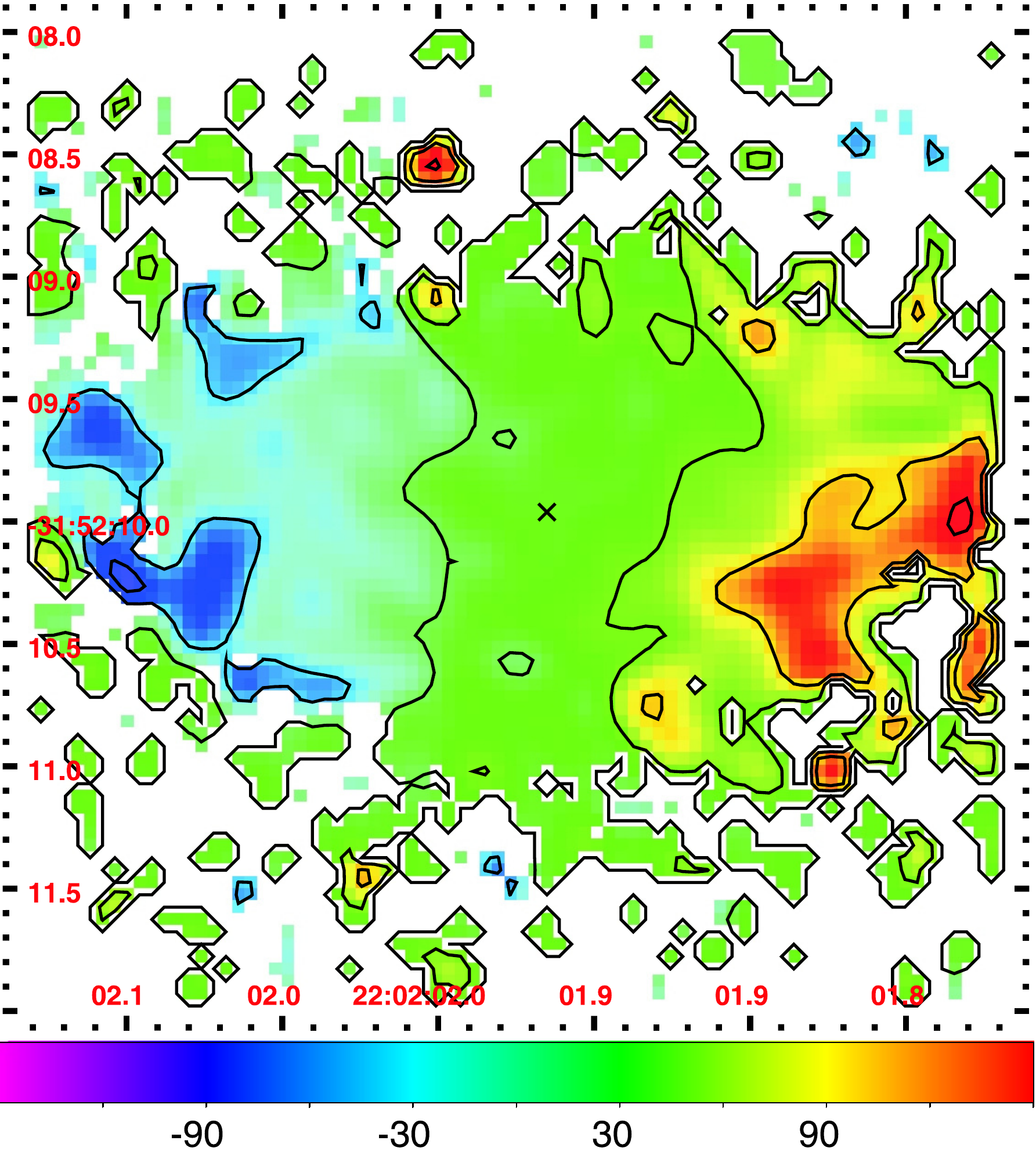}\label{fig:Paavel}}\quad
	\subfigure[SiVI flux]{\includegraphics[width=0.28\textwidth]{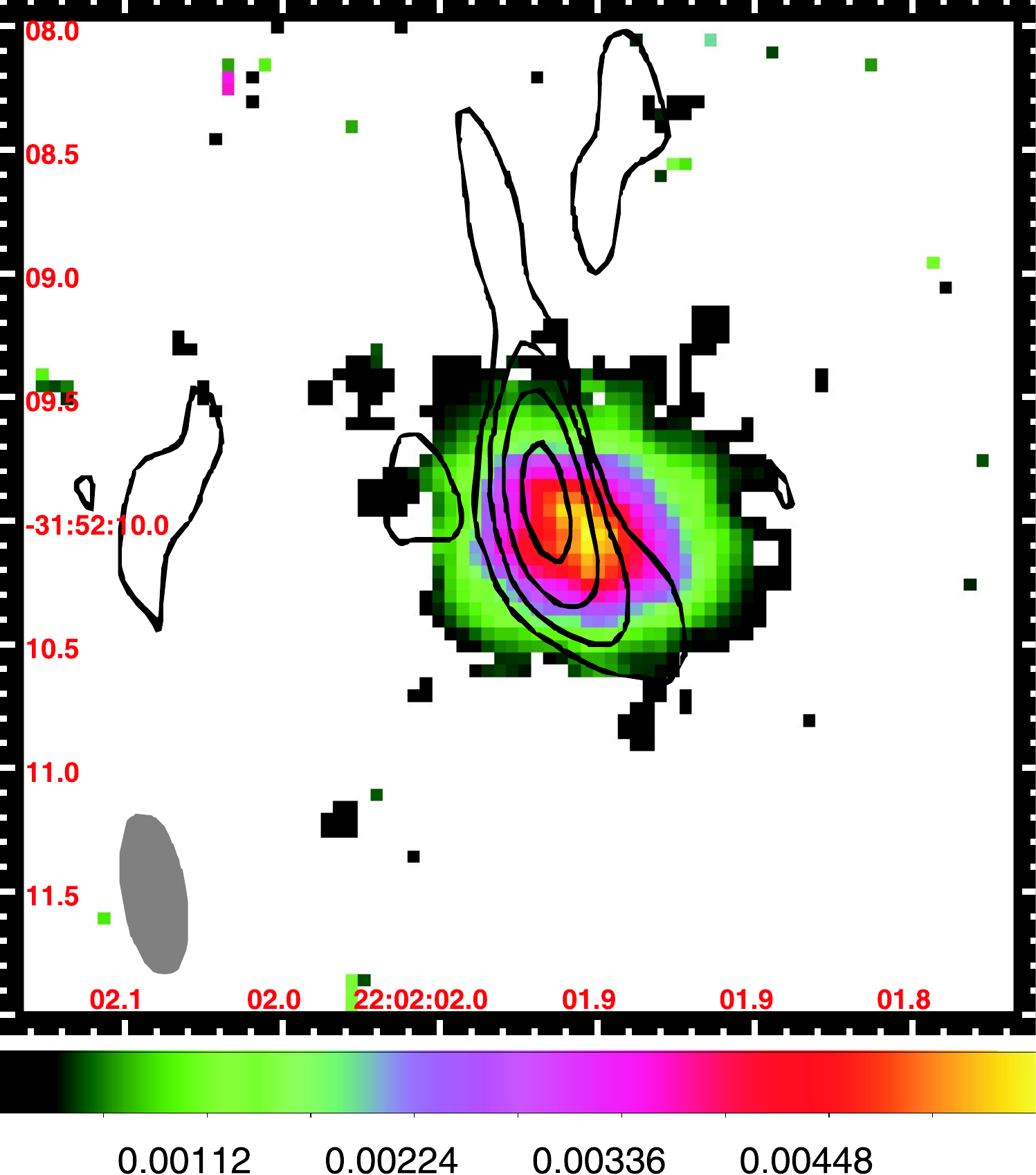}\label{fig:sivifl}}
	\subfigure[H$_2\lambda2.12\mu$m flux]{\includegraphics[width=0.28\textwidth]{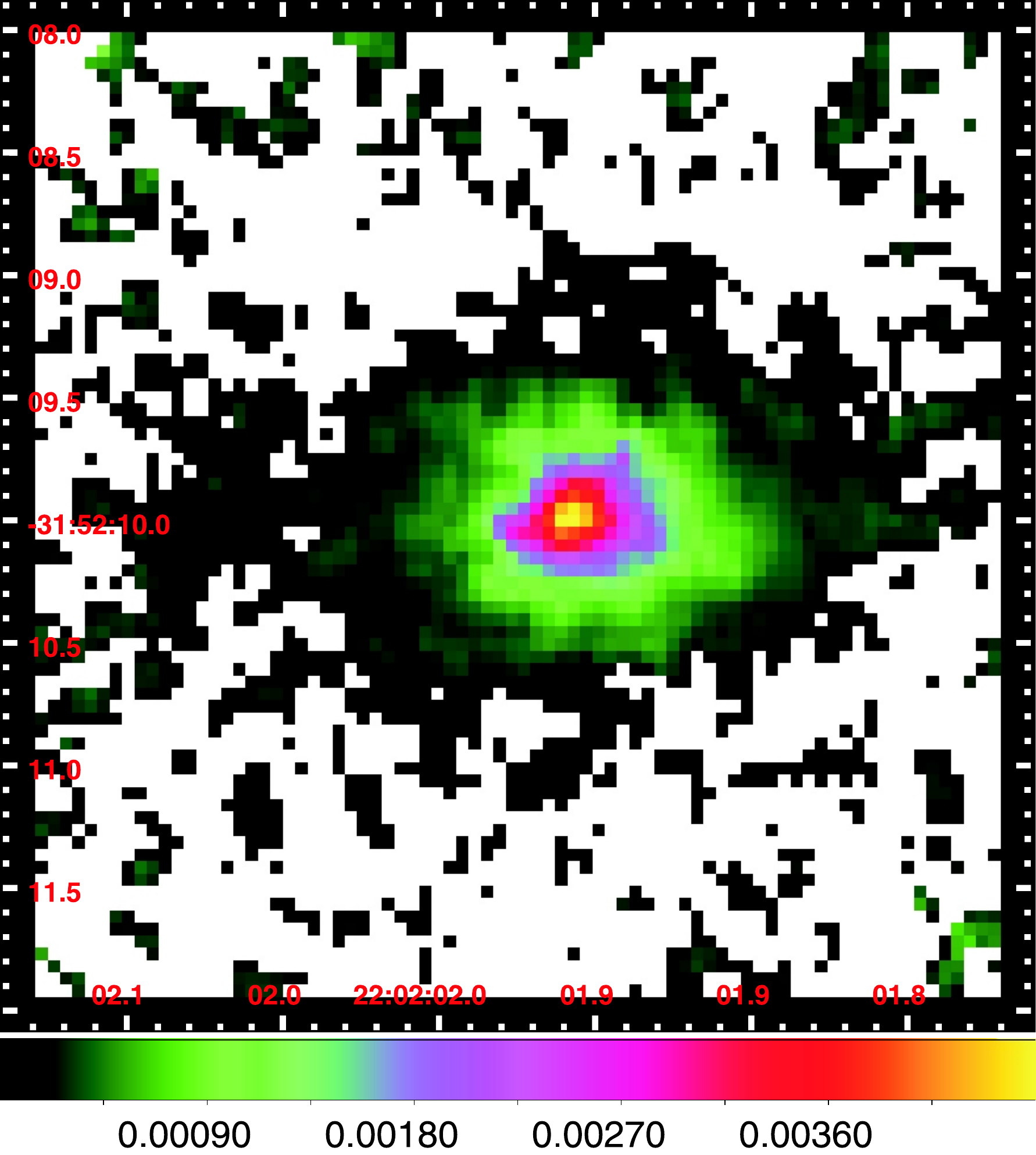}\label{fig:H2fl}}\quad
	\subfigure[H$_2\lambda2.12\mu$m LOSV]{\includegraphics[width=0.28\textwidth]{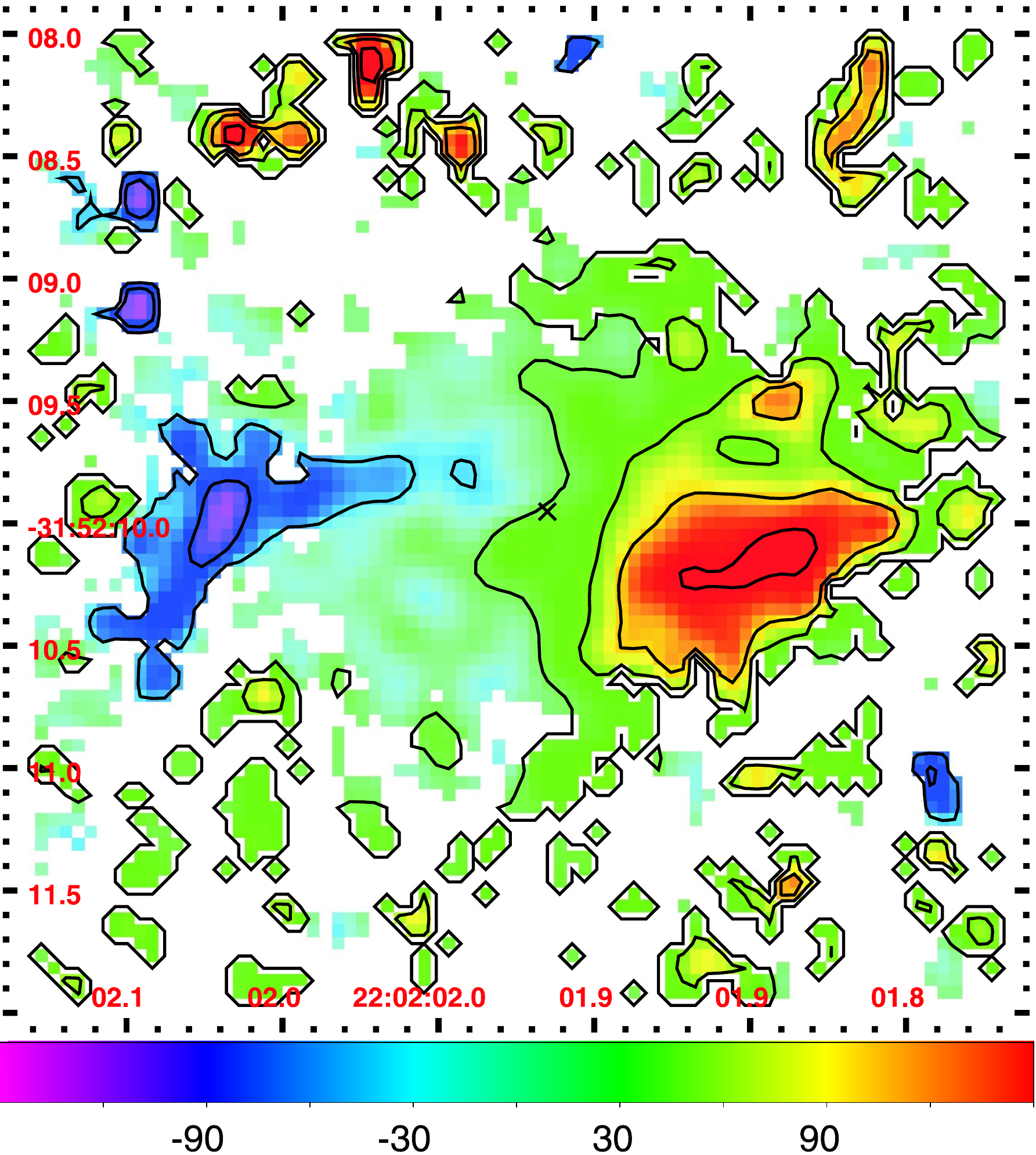}\label{fig:H2vel}}\quad
	\subfigure[SiVI LOSV]{\includegraphics[width=0.28\textwidth]{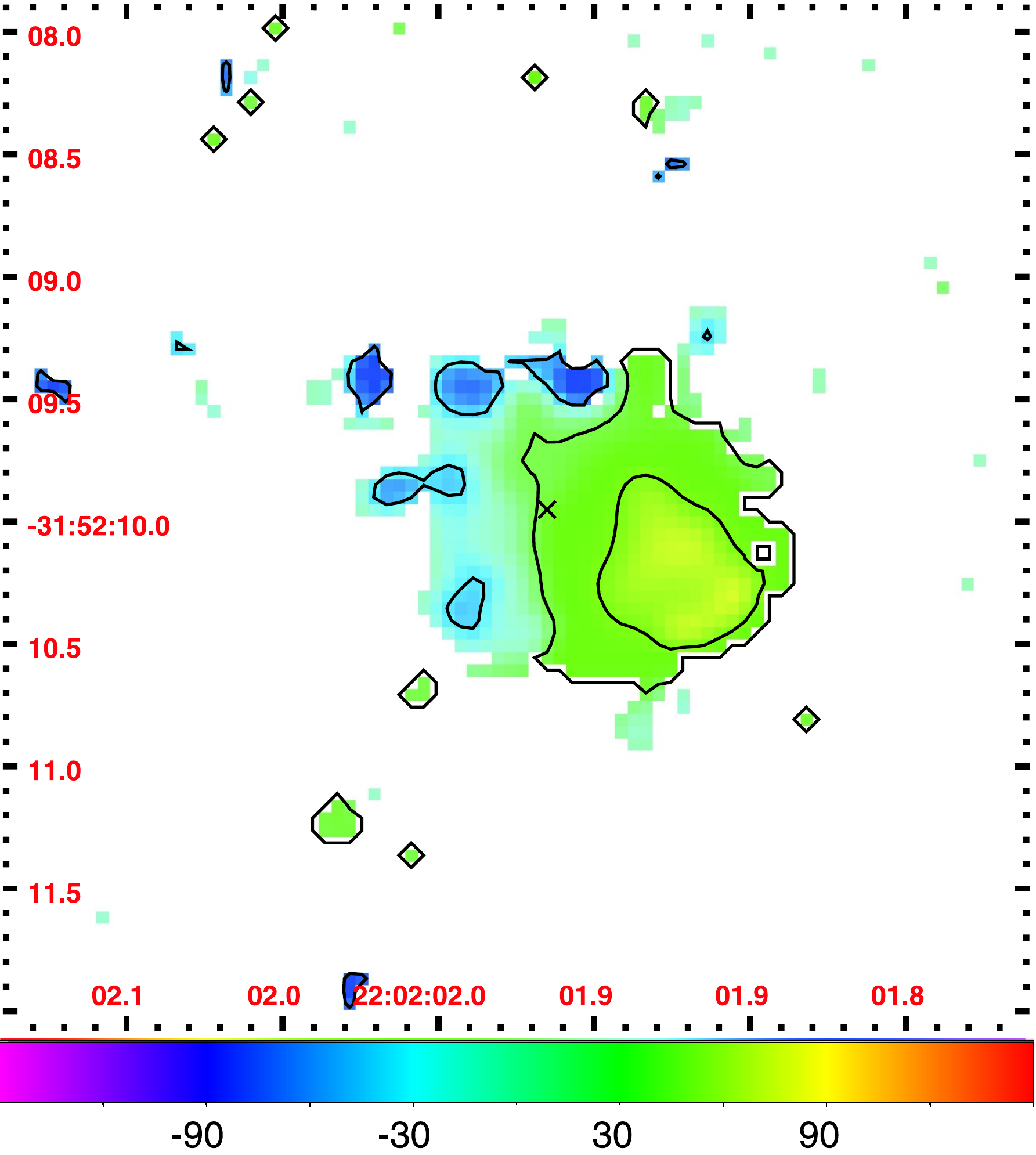}\label{fig:SiVIvel}}
	\subfigure[CO(2-0) LOSV]{\includegraphics[width=0.28\textwidth]{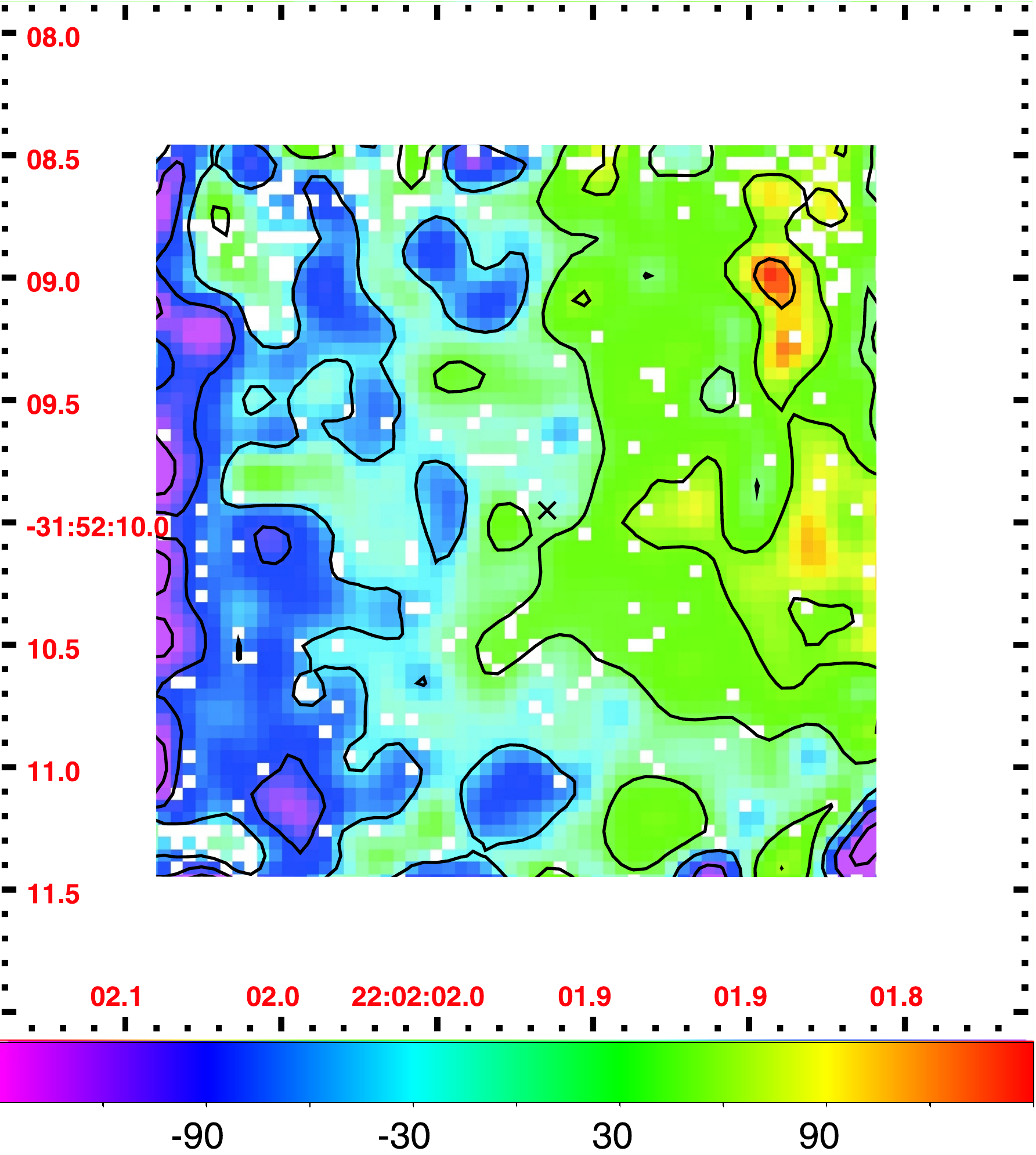}\label{fig:CO20}}\quad
	\subfigure[CO(6-3) LOSV]{\includegraphics[width=0.28\textwidth]{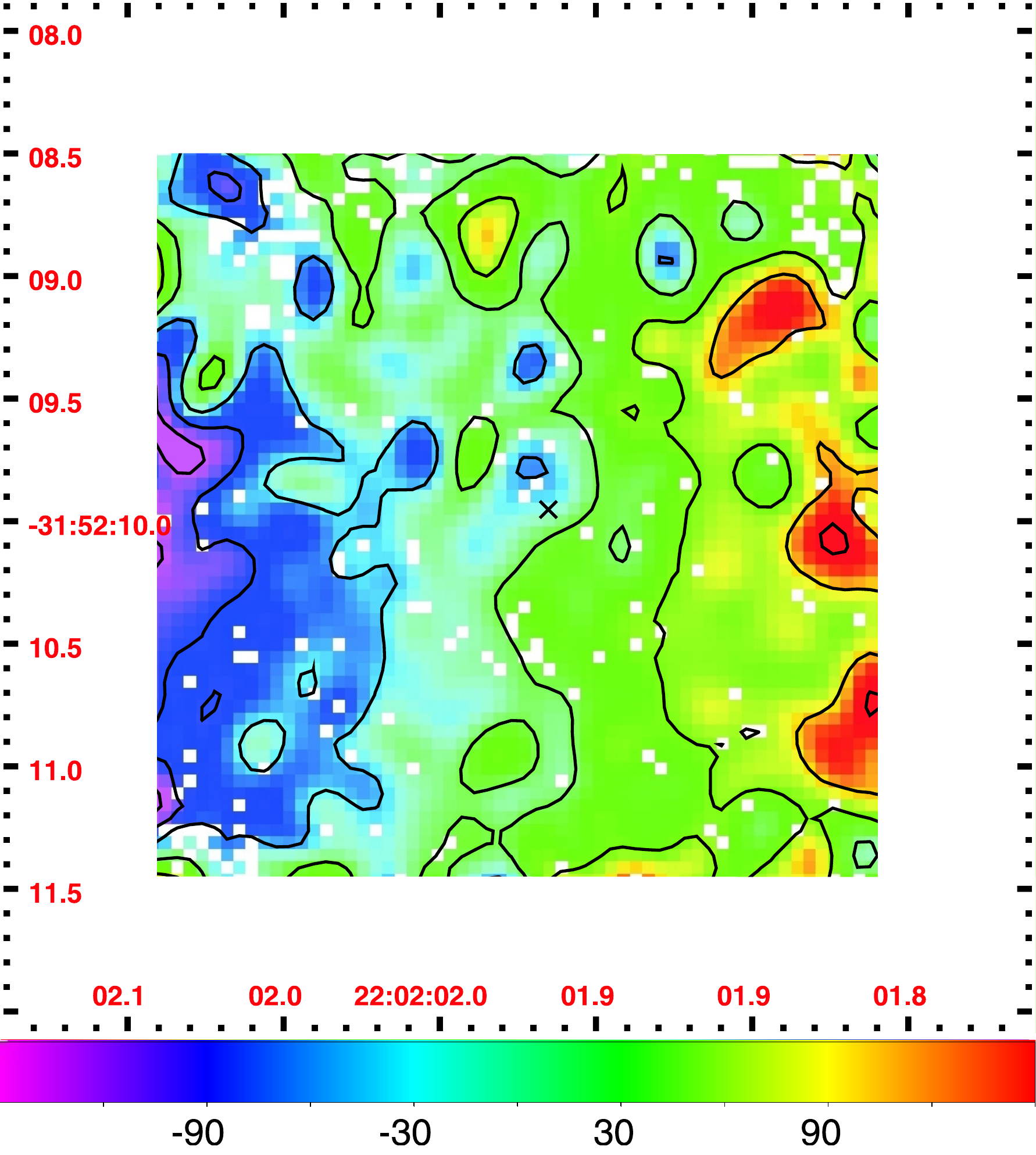}\label{fig:CO63}}\quad
	\subfigure[SiVI LOSVD]{\includegraphics[width=0.28\textwidth]{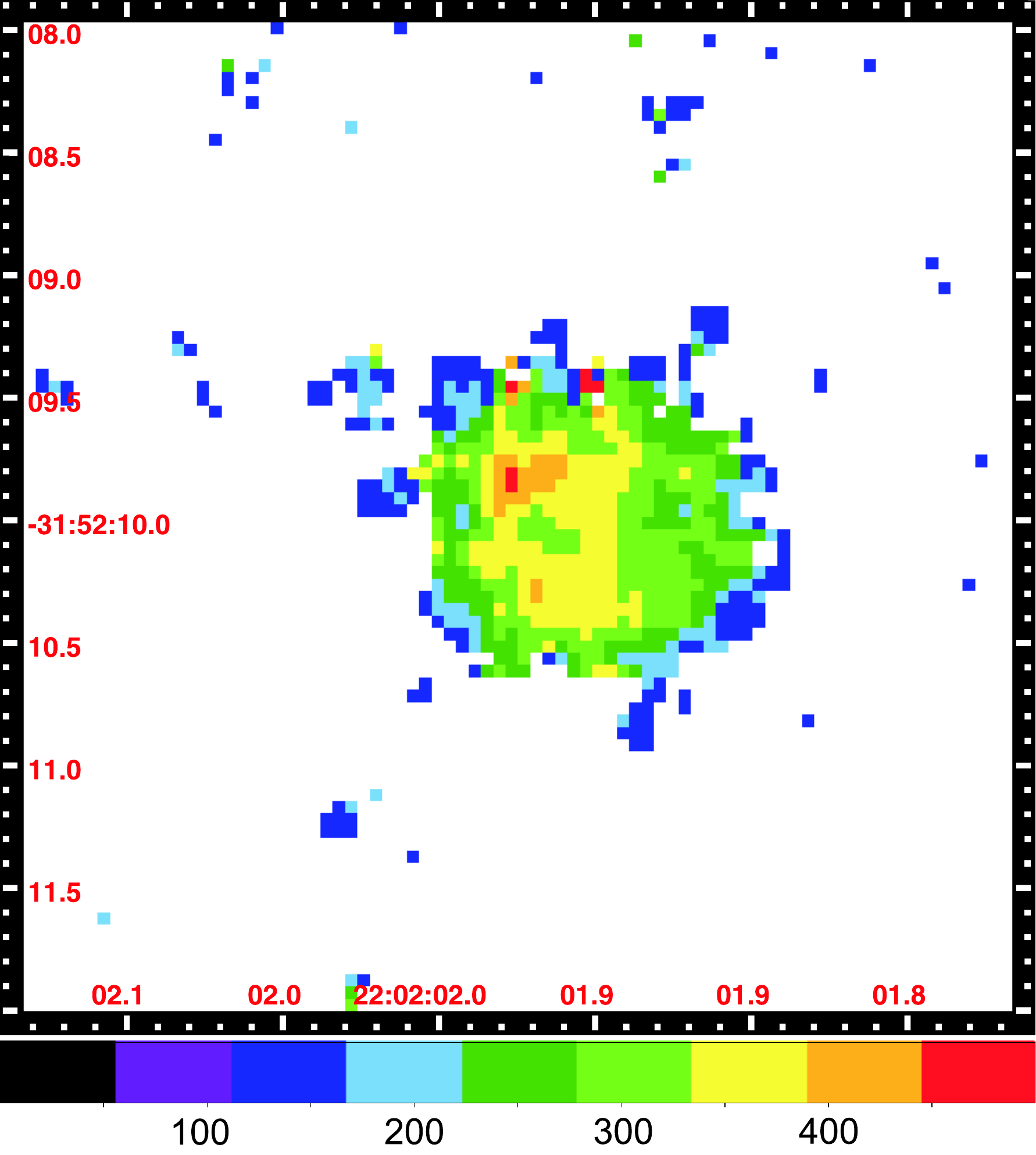}\label{fig:SiVIdisp}}
 \caption{The flux maps are shown in units of 10$^{-16}$Wm$^{-2}\mu$m$^{-1}$, the velocity maps are in units of kms$^{-1}$. Subfigures taken from \cite{2012A&A...544A.105S}. The LOSVD is not corrected for instrumental broadening.
}
 \label{fig:figures}
\end{figure*}

An off nuclear spectrum shows the stellar absorption features that are also visible in the NGC 289 nuclear spectrum (Fig. \ref{fig:nucspec289}). We used these features to determine the LOSV maps (see Fig. \ref{fig:CO20},\subref{fig:CO63}) and the LOSVD in the H- and K-band using the CO(6-3) and CO(2-0) absorption features. Again we see a rotation pattern with blue shift in the east and red shift in the west. The velocities are not equal to the velocities of the emission lines but similar. At the same time we determine the stellar dispersion $\sigma_{\ast}$ from these features. The stellar dispersion enables us to estimate an upper limit for the black hole mass using the measured stellar velocity dispersion of $\sigma_{\ast}\sim$200 km s$^{-1}$. We can compare the upper limit of M$_{BH\downarrow}$=4$\times$10$^{8}$M$_{\odot}$ to our prior calculated BH mass of M$_{BH}$=(4.5$\pm$1.0)$\times$10$^{8}$M$_{\odot}$ and we conclude that these are in good agreement.

\section{Summary and Conclusions}
We presented two similar 'looking' galaxies that, at a closer look, are not that similar. NGC 7172 and NGC 289 are both classified as Seyfert 2 galaxies, however, as NGC 289 shows no characteristics of inhabiting a BLR, NGC 7172 does. NGC 7172 shows a very red nuclear spectrum and broad H-recombination lines. The red spectrum originates from the hot dust of the inner part of the dusty torial structure which is supposed to surround an active SMBH. The broad Pa$\alpha$ and Br$\gamma$ recombination lines are a direct proof of a BLR and the strong hot dust in the center show that NGC 7172 is Seyfert 1 galaxy, according to the UM. Using the broad emission lines and the stellar velocity dispersion in our FOV we are able to determine two independent BH mass estimates for NGC 7172 from out data. Both estimates are in agreement with each other and with literature values. We show that the NIR is far more potent in discovering the true nature of AGN than the optical, even the polarized, light. For more information on NGC 7172 see \cite{2012A&A...544A.105S}.

From this result we can conclude that Seyfert 2 galaxies are intrinsically the same as Seyfert 1 galaxies. The obscuration of NGC 7172 by the dust leads to a misclassification of NGC 7172 as a Seyfert 2 galaxy. Still, the UM proposes that Seyfert 1 and Seyfert 2 galaxies are different to the observer due to the obscuration of the nuclear region by the dusty torus but in the case of NGC 7172 the obscuration takes place in the galactic plane due to the edge-on viewing angle and the thick dust lane crossing the central region of the galaxy. The question that arises is: Are there other galaxies with a similar orientation and amount of dust that are misclassified as Seyfert 2 galaxies but actually are Seyfert 1 galaxies? There is a bias of Seyfert 2 galaxies by Seyfert 1 galaxies though the extent is not clear and a thorough investigation in this field is needed to ensure a statistical independence from orientation and dust coverage. Only then models that want to extend the UM can be proved.


\begin{thebibliography}{99}
 \bibitem{2012A&A...544A.105S} Smaji{\'c}, S., Fischer, S., Zuther, J., \& Eckart, A.\ 2012, \aap, 544, A105 
 \bibitem{2008ARA&A..46..475H} Ho, L.~C.\ 2008, \araa, 46, 475 
 \bibitem{2007MNRAS.382.1415S} Schawinski, K., 
Thomas, D., Sarzi, M., et al.\ 2007, \mnras, 382, 1415 
 \bibitem{1993ARA&A..31..473A} Antonucci, R.\ 1993, \araa, 31, 473 
 \bibitem{2006ApJ...653..137Z} Zhang, E.-P., \& Wang, J.-M.\ 2006, \apj, 653, 137
 \bibitem{2013PoS...M.V-S.} Valencia-S., M. et al., this proceedings
 \bibitem{2012A&AT...27..557A} Antonucci, R.\ 2012, Astronomical and Astrophysical Transactions, 27, 557 
 \bibitem{2010ApJ...725L.210S} Shen, S., Shao, Z., 
\& Gu, M.\ 2010, \apjl, 725, L210 
 \bibitem{2011MNRAS.414.2148L} Lagos, C.~D.~P., Padilla, 
N.~D., Strauss, M.~A., Cora, S.~A., \& Hao, L.\ 2011, \mnras, 414, 2148 
 \bibitem{2004ApJ...616..707H} Hunt, L.~K., \& Malkan, M.~A.\ 2004, \apj, 616, 707 
 \bibitem{2006A&A...455..773V} V{\'e}ron-Cetty, M.-P., \& V{\'e}ron, P.\ 2006, \aap, 455, 773
 \bibitem{2003SPIE.4841.1548E} Eisenhauer, F., 
Abuter, R., Bickert, K., et al.\ 2003, \procspie, 4841, 1548
 \bibitem{1979ApJS...40..657M} Marshall, F.~E., 
Boldt, E.~A., Holt, S.~S., et al.\ 1979, \apjs, 40, 657 
 \bibitem{2001MNRAS.327..459L} Lumsden, S.~L., 
Heisler, C.~A., Bailey, J.~A., Hough, J.~H., 
\& Young, S.\ 2001, \mnras, 327, 459 
\bibitem{1998MNRAS.298..824G} Guainazzi, M., Matt, 
G., Antonelli, L.~A., et al.\ 1998, \mnras, 298, 824 
 \bibitem{2002AJ....123..159C} Carollo, C.~M., 
Stiavelli, M., Seigar, M., de Zeeuw, P.~T., 
\& Dejonghe, H.\ 2002, \aj, 123, 159 
 \bibitem{2004AJ....127.3338B} Bendo, G.~J., \& Joseph, R.~D.\ 2004, \aj, 127, 3338 
 \bibitem{2010ApJ...724..386K} Kim, D., Im, M., 
\& Kim, M.\ 2010, \apj, 724, 386 
 \bibitem{1994A&A...282..418L} Longo, G., Busarello, G., Lorenz, H., Richter, G., \& Zaggia, S.\ 1994, \aap, 282, 418 
 \bibitem{2006ApJ...645..928A} Awaki, H., Murakami, H., 
Ogawa, Y., \& Leighly, K.~M.\ 2006, \apj, 645, 928 
 \bibitem{2009ApJ...690.1322W} Winter, L.~M., 
Mushotzky, R.~F., Reynolds, C.~S., \& Tueller, J.\ 2009, \apj, 690, 1322 
\end{thebibliography}
\end{document}